\documentclass[10pt,twocolumn]{article}

\usepackage[T1]{fontenc}
\usepackage{lmodern} 

\usepackage{amsmath,amssymb,amsthm}

\usepackage{graphicx}
\usepackage{float}
\usepackage{booktabs}
\usepackage{multirow}
\usepackage{tabularx}
\usepackage{array}
\newcolumntype{Y}{>{\raggedright\arraybackslash}X}
\newcolumntype{Z}{>{\centering\arraybackslash}X}
\newcolumntype{W}{>{\raggedleft\arraybackslash}X}
\renewcommand{\arraystretch}{1.3}

\usepackage{subcaption}

\usepackage{enumitem}
\setlist{topsep=3pt, itemsep=3pt, parsep=0pt, partopsep=0pt}

\setlist[enumerate,1]{label=\alph*), leftmargin=*}

\usepackage[numbers,sort&compress]{natbib}

\usepackage[colorlinks=true,
linkcolor=blue,
citecolor=blue,
urlcolor=blue]{hyperref}

\usepackage{geometry}
\newtheorem{theorem}{Theorem}

\newtheorem{proposition}{Proposition}

\usepackage{titlesec}
\titleformat{\section}
{\normalsize\bfseries}
{\thesection.}
{0.5em}
{}
\titlespacing*{\section}{0pt}{2.5ex plus 0.2ex minus 0.2ex}{1.0ex}

\titleformat{\subsection}
{\normalsize\itshape}
{\thesubsection}
{0.5em}
{}
\titlespacing*{\subsection}{0pt}{1.5ex plus 0.2ex minus 0.2ex}{0.8ex}

\titleformat{\subsubsection}
{\normalsize\itshape}
{\thesubsubsection}
{0.6em}
{}
\titlespacing*{\subsubsection}{0pt}{1.2ex plus 0.2ex minus 0.2ex}{0.6ex}

\titleformat{\paragraph}
{\normalsize\bfseries}
{}
{0em}
{}

\titlespacing*{\paragraph}
{0pt}        
{1.2ex}      
{0.2em}      

\usepackage[font=small,labelfont=bf]{caption}
\DeclareMathOperator{\Var}{Var}

\DeclareMathOperator{\I}{I}

\newcommand{\USL}{\mathrm{USL}}
\newcommand{\LSL}{\mathrm{LSL}}

\usepackage[dvipsnames]{xcolor}
\usepackage[normalem]{ulem}
\usepackage{soul}
\sethlcolor{yellow!35}

\soulregister\cite7
\soulregister\citet7
\soulregister\citep7
\soulregister\ref7
\soulregister\pageref7

\usepackage{framed}
\definecolor{shadecolor}{rgb}{1,1,0.65}
\definecolor{mygrey}{gray}{0.965}

\title{Finite-Sample Decision Instability in Threshold-Based Process Capability Approval}

\author{
	Fei Jiang\thanks{Independent Researchers, Seattle, WA, USA. Corresponding author: Fei Jiang (jiangfeicq@gmail.com).}
	\and
	Lei Yang\footnotemark[1]
}

\date{} 
\begin{document}
\maketitle
\begin{abstract}
	Process capability indices such as $C_{pk}$ are widely used in manufacturing quality control to support supplier qualification and product release decisions based on fixed acceptance thresholds (e.g., $C_{pk} \geq 1.33$). In practice, these decisions rely on sample-based estimates computed from moderate sample sizes ($n \approx$ 20-50), yet the stochastic nature of the estimator is often overlooked when interpreting threshold compliance. This study establishes a local asymptotic characterization of decision behavior when the true process capability lies near a fixed threshold. Under standard regularity conditions, if the true capability equals the threshold, the acceptance probability converges to 0.5 as sample size increases, implying that a fixed $C_{pk}$ gate embeds an inherent boundary decision risk even under ideal distributional assumptions. When the true capability deviates from the threshold by $O(n^{-1/2})$, the decision probability converges to a non-degenerate limit governed by a scaled signal-to-noise ratio. Monte Carlo simulations and an empirical study on 880 manufacturing dimensions demonstrate substantial resampling-based decision instability near the commonly used 1.33 criterion. These findings provide a probabilistic interpretation of threshold-based capability decisions and quantitative guidance for assessing boundary-induced release risk in engineering practice.
\end{abstract}


\setcounter{tocdepth}{2}  
\section{Introduction}

Process capability indices (PCIs) such as $C_p$, $C_{pk}$, $P_p$, and $P_{pk}$ are foundational tools in manufacturing quality engineering for assessing whether a process can consistently operate within specification limits \cite{juran1979quality, kane1986process, kotz2002process}. 
These indices are standardized in ISO/TR 22514-4 \cite{ISO22514-4-2016} and embedded within broader statistical process control (SPC) frameworks \cite{ISO11462-1-2001, ISO11462-2-2010}. 
In industrial practice, capability thresholds such as $1.00$, $1.33$, and $1.67$ are routinely adopted as deterministic acceptance criteria for supplier qualification, pilot production, and release decisions. 
Operationally, capability evaluation therefore functions as a binary decision relative to a predefined threshold.

Although widely interpreted as deterministic quality metrics, PCIs are statistical estimators computed from finite samples. 
In many practical settings, such as engineering builds, dimensional audits, and early-stage production runs, the available sample size is limited. 
Consequently, both the sample mean and standard deviation are random, and the estimated index $\widehat{C}_{pk}$ is itself random. When approval decisions are based directly on the rule $\widehat{C}_{pk} \ge C_0$, where $C_0$ denotes a fixed approval threshold, sampling variability induces non-negligible misclassification risk. 
Related misclassification phenomena have been studied in conformity assessment and measurement system analysis \cite{burdick2005confidence, zappa2009misclassification}, but those works focus on observation-level measurement uncertainty rather than estimator-induced variability in capability indices.

The existing PCI literature has extensively addressed point estimation, interval estimation, and hypothesis testing procedures for capability parameters \cite{mathew2007generalized, pearn1992distributional}. 
Lower confidence bounds (LCBs) and confidence-interval-based release rules are widely used to account for estimator uncertainty in capability assessment \cite{kushler1992confidence, grau2011lower}. Early approximate confidence limits for $C_{pk}$ were developed to form the basis of subsequent LCB-based qualification procedures \cite{zhang1990interval}. These methods provide inferential guarantees for the unknown capability parameter, but do not explicitly characterize the stochastic reliability of the fixed threshold rule $\I(\widehat C_{pk} \ge C_0)$. While such approaches provide inferential guarantees for the unknown parameter $C_{pk}$, they do not explicitly characterize the reliability of the induced threshold-based decision rule itself. 
In particular, the probability that the deterministic rule yields a correct classification has not been analyzed as an explicit function of true capability and sample size.

Unlike hypothesis testing procedures that calibrate rejection regions to achieve a prespecified Type I error rate (i.e., a fixed probability of falsely rejecting a true null hypothesis), industrial threshold rules are fixed ex ante and do not adjust with sample size \cite{casella2024statistical}. Consequently, the induced misclassification risk is not controlled but instead emerges endogenously from estimator dispersion. The boundary-instability phenomenon is therefore structurally distinct from classical size/power analysis. Unlike classical hypothesis tests that calibrate rejection regions to control Type I error, capability thresholds implicitly embed uncontrolled decision risk.

To our knowledge, while the inferential properties of capability estimators have been extensively studied, the stochastic reliability of the deterministic threshold rule itself has not been formally characterized \cite{mathew2007generalized}. In particular, the limiting acceptance probability at the approval boundary has not been explicitly derived in the existing PCI literature.

This paper reformulates capability approval as a statistical decision problem. Specifically, we consider the conventional threshold-based rule $D = \I(\widehat{C}_{pk} \ge C_0)$, which represents the deterministic release criterion used in practice. Under finite samples, however, $D$ is random because $\widehat{C}_{pk}$ is random. We therefore quantify decision reliability through the conditional misclassification probabilities induced by the sampling distribution of $\widehat{C}_{pk}$.

Under standard regularity conditions ensuring asymptotic normality of smooth estimators \cite{lehmann1998theory, van2000asymptotic, casella2024statistical}, and provided that the minimum defining $C_{pk} = \min(C_{pu}, C_{pl})$ is uniquely attained at the true parameter, we show that when $C_{pk}^{true} = C_0$,
\[
\Pr(\widehat{C}_{pk} \ge C_0) \to 0.5
\quad \text{as } n \to \infty.
\]
This boundary behavior reflects symmetry of the limiting distribution at the decision threshold and implies boundary instability at the approval threshold.

In industrial settings, capability thresholds often function as contractual or regulatory quality gates. Processes operating near these thresholds may be economically optimized to balance yield and quality risk. In such regimes, even moderate sampling variability can materially alter release outcomes. Understanding the probabilistic stability of such threshold-based decisions is therefore not merely theoretical but directly relevant to manufacturing governance and supplier management.

Recent work has proposed practical workflows for computing and interpreting process capability indices in industrial settings \cite{jiang2026practical}. Building on these operational perspectives, the present study focuses on the statistical reliability of the threshold-based approval decisions that such workflows ultimately support.

This motivates the central research question of the present study: How reliable is the conventional deterministic capability threshold rule when capability indices are estimated from finite samples?

To address this question, we develop a statistical framework that characterizes the reliability of threshold-based capability
approval decisions under finite-sample uncertainty. The main contributions of this paper are:

\begin{enumerate}
	\item \textbf{Formal decision-theoretic formulation of capability approval:}
	We formalize threshold-based capability qualification as a stochastic classification problem and define misclassification probability as the fundamental measure of decision reliability.
	
	\item \textbf{Boundary instability characterization:}
	We establish a boundary-instability result showing that acceptance probability converges to $0.5$ when the true capability equals the threshold, and we derive the associated signal-to-noise scaling governing finite-sample behavior.
	
	\item \textbf{Finite-sample risk surfaces and empirical relevance:}
	Using Monte Carlo simulation, we construct misclassification probability surfaces that quantify decision instability as a function of true capability and sample size. Analysis of a 880-dimension manufacturing dataset demonstrates that near-threshold operating regimes are common in practice.
\end{enumerate}

By shifting focus from parameter estimation accuracy toward explicit characterization of decision reliability, this work clarifies the probabilistic structure underlying threshold-based capability approval and provides a principled statistical basis for interpreting release decisions under finite-sample uncertainty.

The remainder of the paper develops the statistical foundations of capability estimators (Section~\ref{sec2}), formulates the decision-theoretic framework (Section~\ref{sec3}), establishes the boundary-instability result, and presents finite-sample and empirical analyses (Section~\ref{sec4}).

\section{Sampling Framework}
\label{sec2}

We begin by formalizing the stochastic structure of capability estimation under normal sampling, which serves as the basis for subsequent decision-theoretic analysis.

\subsection{Sampling Variability Under Normal Assumptions}
\label{sec2-a}

Assuming a statistically stable process with normally distributed output and bilateral specifications (not necessarily symmetric), the classical process capability index is defined as \cite{kane1986process, kotz2002process, ISO22514-4-2016}
\begin{equation}
	C_{pk} = C_{pk}(\mu, \sigma)= \min\left(\frac{\USL - \mu}{3\sigma}, \frac{\mu - \LSL}{3\sigma}\right),
	\label{cpk1}
\end{equation}
where $\USL$ and $\LSL$ denote the upper and lower specification limits, and $\mu$ and $\sigma$ denote the true process mean and standard deviation. 
In practice, $\mu$ and $\sigma$ are replaced by their sample estimators $\bar{X}$ and $S$:
\begin{equation}
	\widehat{C}_{pk} = \widehat{C}_{pk}(\bar{X}, S) = \min\left(\frac{\USL - \bar{X}}{3S}, \frac{\bar{X} - \LSL}{3S}\right).
	\label{cpk2}
\end{equation}

Under normal sampling,
\begin{equation}
	\bar{X} \sim \mathrm{N}\!\left(\mu, \frac{\sigma^2}{n}\right),
	\quad
	\frac{(n-1)S^2}{\sigma^2} \sim \chi^2_{n-1}.
\end{equation}

Hence, under finite samples, $\widehat{C}_{pk}$ is a random variable induced by the joint sampling variability of $\bar{X}$ and $S$. Consequently, any decision rule based on thresholding $\widehat{C}_{pk}$ must be understood as a stochastic classification procedure.

We focus on the classical $C_{pk}$ definition used in routine release gates; target-based capability indices (e.g., $C_{pm}$, $C_{pmk}$) are not considered here, as the decision-instability phenomenon arises from the thresholding mechanism rather than from the specific form of the capability functional.

\section{Decision-Theoretic Formulation of Capability Classification}
\label{sec3}
Capability approval based on a threshold rule of the form $\hat{C}_{pk} \ge C_0$ can be formulated as a statistical decision problem in the sense of classical statistical decision theory \cite{berger2013statistical}. Although implemented as a deterministic comparison, the rule induces a stochastic classification mechanism because $\hat{C}_{pk}$ is a random variable under finite sampling.

Within this framework, uncertainty is quantified through the conditional misclassification probabilities induced by the sampling distribution of $\widehat{C}_{pk}$. The decision rule is formally defined as
\begin{equation}
	D = \I(\widehat{C}_{pk} \geq C_0),
	\label{eq:decision_rule}
\end{equation}
where $\I(\cdot)$ denotes the indicator function. Here $D=1$ corresponds to acceptance and $D=0$ to rejection.

To formalize this setting, we distinguish between the true (but unobservable) population capability $C_{pk}^{true}$ and its estimator $\widehat{C}_{pk}$. Misclassification arises whenever the decision based on $\widehat{C}_{pk}$ disagrees with the true process status relative to the threshold $C_0$.

\subsection{Misclassification Formulation}
\label{sec3-a}

\subsubsection{Type I Misclassification (False Accept)}
\begin{equation}
	{\Pr}_{Type\ I} = \Pr(\widehat{C}_{pk} \ge C_0 \mid C_{pk}^{true} < C_0).
\end{equation}
This probability measures the likelihood of accepting a process whose true capability lies below the threshold.

\subsubsection{Type II Misclassification (False Reject)}
\begin{equation}
	{\Pr}_{Type\ II} = \Pr(\widehat{C}_{pk} < C_0 \mid C_{pk}^{true} \ge C_0).
\end{equation}
This probability measures the likelihood of rejecting a process whose true capability meets or exceeds the threshold.

Note that, unlike classical hypothesis testing frameworks in which the Type I error is typically fixed at a prespecified level, here both Type I and Type II misclassification probabilities vary continuously with the true capability level and the sample size. The threshold rule therefore induces a parameter-dependent decision risk rather than a controlled testing procedure.

These error probabilities characterize the stochastic behavior of threshold-based capability classification as a function of the true underlying capability level.

\subsection{Finite-Sample Decision Performance}
\label{sec3-b}

Under finite sampling, the threshold rule defined in Eq. \eqref{eq:decision_rule} induces a nondegenerate risk function. Even when the true capability $C_{pk}^{true}$ lies strictly on one side of the threshold, the sampling distribution of $\widehat{C}_{pk}$ assigns positive probability to the opposite classification.

The rule is formally equivalent to a one-sided testing problem with $H_0: C_{pk}^{true} < C_0$ and $H_1: C_{pk}^{true} \ge C_0$, in which the misclassification probabilities coincide with the corresponding Type I and Type II error rates \cite{pearn2004testing}.

Unlike classical testing frameworks that control Type I error at a prespecified level, the present analysis characterizes the unconditional acceptance probability induced by deterministic industrial thresholds, which are typically fixed by contractual or regulatory standards rather than statistical design.

The misclassification probabilities depend smoothly on the location of $C_{pk}^{true}$ relative to the threshold. They attain their largest values in neighborhoods of the decision boundary $C_{pk}^{true} = C_0$, where the sampling distribution of $\widehat{C}_{pk}$ overlaps both acceptance and rejection regions.

Accordingly, the finite-sample decision performance is fully characterized by the pair of conditional misclassification probabilities as functions of $(C_{pk}^{true}, n)$.

\subsection{Boundary Instability Theorem}
\label{sec3-c}
Now characterize the asymptotic behavior of the decision probability under the threshold rule defined in Eq. \eqref{eq:decision_rule}.

\paragraph{Assumption A1 (Unique Active Constraint).}
The true parameter satisfies $(\USL - \mu) \neq (\mu - \LSL)$, so that in the definition of $C_{pk}(\mu,\sigma)$ given in Eq.~\eqref{cpk1} the minimum is uniquely attained at the true parameter. Hence the capability functional is locally differentiable in a neighborhood of $(\mu,\sigma)$.
 
\paragraph{Assumption A2 (Regularity).}
The estimators $(\bar{X}, S)$ admit the joint asymptotic normal expansion
\begin{equation}
	\label{eq:asymptotic_joint}
	\sqrt{n}
	\begin{pmatrix}
		\bar{X} - \mu \\
		S - \sigma
	\end{pmatrix}
	\xrightarrow{d}
	\mathcal{N}(0,\Sigma),
\end{equation}
where $\Sigma$ is positive definite.

Under Assumptions A1-A2, the multivariate delta method \cite{van2000asymptotic} yields
\begin{equation}
	\sqrt{n}(\widehat{C}_{pk} - C_{pk}^{true})
	\xrightarrow{d}
	\mathcal{N}(0,\sigma_C^2),
	\label{sigmac}
\end{equation}
where the asymptotic variance is given by
\begin{equation}
	\label{sigmac2}
	\sigma_C^2
	=
	\nabla g(\mu,\sigma)^\top
	\Sigma
	\nabla g(\mu,\sigma).
\end{equation}

The magnitude of $\sigma_C$ depends on both process centering and dispersion, as reflected in the gradient $\nabla g(\mu,\sigma)$, and on the joint variability of the estimators $(\bar{X}, S)$ through the covariance matrix $\Sigma$. Consequently, decision sensitivity near the approval boundary is influenced not only by sample size but also by specification geometry and operating conditions. Equivalently,
\[
\widehat{C}_{pk}
=
C_{pk}^{true}
+
\frac{\sigma_C}{\sqrt{n}} Z
+
o_p(n^{-1/2}),
\quad
Z \sim \mathcal{N}(0,1),
\]

where $o_p(n^{-1/2})$ denotes a remainder term that is
asymptotically negligible relative to $n^{-1/2}$ in probability.

\begin{theorem}[Local Boundary Instability]
	\label{thm:boundary}
	
	Let $C_0$ be a fixed threshold. Under Assumptions A1-A2,
	
	\begin{equation}
		\label{eq:asymp_accept}
		\Pr\!\bigl(\widehat{C}_{pk} \ge C_0\bigr)
		=
		\Phi\!\left(
		\frac{\sqrt{n}\,(C_{pk}^{true}-C_0)}{\sigma_C}
		\right)
		+ o(1),
		\ n \to \infty,
	\end{equation}
	where $\Phi(\cdot)$ is the standard normal distribution function.
	
	Moreover, consider a local parameterization
	\begin{equation}
		\label{cpkh}
		C_{pk}^{true}
		=
		C_0
		+
		\frac{h}{\sqrt{n}},
		\quad h \in \mathbb{R},
	\end{equation}
	where $h=\sqrt{n}(C_{pk}^{true}-C_0)$
	is the $\sqrt{n}$-scaled distance of the true capability level from the
	approval threshold. Then
	\[
	\Pr\!\bigl(\widehat{C}_{pk} \ge C_0\bigr)
	\;\rightarrow\;
	\Phi\!\left(\frac{h}{\sigma_C}\right).
	\]
	
	In particular, at the exact boundary $C_{pk}^{true} = C_0$,
	\[
	\lim_{n\to\infty}
	\Pr\!\bigl(\widehat{C}_{pk} \ge C_0\bigr)
	=
	\frac{1}{2}.
	\]
	
\end{theorem}

Theorem \ref{thm:boundary} formalizes boundary instability in deterministic threshold-based capability approval. When the true capability equals the approval threshold, no increase in sample size can eliminate the limiting 50\% acceptance probability. This phenomenon is therefore not a small-sample artifact but a direct consequence of combining a $\sqrt{n}$-consistent estimator with a fixed classification threshold.

Intuitively, when the true capability equals the approval threshold, the sampling distribution of the estimator is asymptotically symmetric around the boundary. As a result, half of repeated samples fall above the threshold and half below, producing the limiting 50\% acceptance probability.

Moreover, the theorem shows that decision behavior near the threshold is governed by the signal-to-noise ratio $\sqrt{n}(C_{pk}^{true}-C_0)/\sigma_C$. Decision instability persists whenever the true capability lies within approximately
$\sigma_C/\sqrt{n}$ of the threshold, so that the sampling distribution places substantial probability mass on both sides of the decision boundary. 

More generally, the result shows that deterministic capability thresholds generate a non-vanishing decision uncertainty region of width proportional to $1/sqrt{n}$, which can’t be eliminated by increasing sample size unless the true capability departs from the threshold at a faster rate.

\begin{proposition}[Explicit $1/\sqrt{n}$ width of the instability region]
	\label{prop:instability_width}
	Assume the regular regime of Theorem~\ref{thm:boundary}, so that the acceptance probability satisfies the asymptotic expansion in Eq. \eqref{eq:asymp_accept}. Fix any $\varepsilon \in (0,1/2)$ and define the $\varepsilon$-instability band
	\[
	\mathcal{B}_{n}(\varepsilon)
	:=
	\left\{
	C_{pk}^{true} :
	\left|
	\Pr(\widehat{C}_{pk} \ge C_0) - \frac{1}{2}
	\right|
	\le \varepsilon
	\right\}.
	\]

	Then, as $n\to\infty$, with $z_\varepsilon := \Phi^{-1}\!\left(\tfrac12+\varepsilon\right)$,
	\[
	\mathcal{B}_{n}(\varepsilon)
	=
	\left[
	C_0 - \frac{\sigma_C z_\varepsilon}{\sqrt{n}},
	\;
	C_0 + \frac{\sigma_C z_\varepsilon}{\sqrt{n}}
	\right]
	+ o\!\left(n^{-1/2}\right).
	\]
	
	and therefore the width of the instability band satisfies
	\[
	\mathrm{width}\bigl(\mathcal{B}_{n}(\varepsilon)\bigr)
	=
	\frac{2\sigma_C}{\sqrt{n}}\,
	\Phi^{-1}\!\left(\frac12+\varepsilon\right)
	+ o\!\left(\frac{1}{\sqrt{n}}\right).
	\]
	In particular, the decision-sensitive region around the approval threshold contracts at the canonical $1/\sqrt{n}$ rate.
\end{proposition}

\paragraph{Visualization remark.}
For visualization purposes, the boundary-sensitive region may be illustrated using the canonical $1/\sqrt{n}$ contraction rate implied by the asymptotic expansion above, without imposing a specific closed-form expression for $\sigma_C$.

\paragraph{Proof sketch.}
Under Assumptions A1--A2, the delta method yields the asymptotic
expansion in Eq.~\eqref{sigmac}, implying that
$\sqrt{n}(\widehat{C}_{pk}-C_{pk}^{true})$ is asymptotically
$\mathcal{N}(0,\sigma_C^2)$. Hence,
\[
\Pr(\widehat{C}_{pk} \ge C_0)
=
\Pr\!\left(
Z \ge
\frac{\sqrt{n}(C_0-C_{pk}^{true})}{\sigma_C}
\right)
+ o(1),
\]
where $Z\sim N(0,1)$.
Substituting the local parameterization
$C_{pk}^{true}=C_0+h/\sqrt{n}$
yields the stated limit.

\subsubsection{Closed-Form Approximation of $\sigma_C$ under Normality}

To operationalize the boundary-instability result,
we derive an explicit approximation for $\sigma_C$
under the standard normal-process assumption.

Assume $X_1,\dots,X_n \sim \mathcal N(\mu,\sigma^2)$
and that Assumption~A1 holds (i.e., exactly one specification
side is active). Without loss of generality,
suppose the upper specification limit is active.
Then the capability functional can be written locally as
\begin{equation}
	\label{cpku}
	C_{pk}^{true} = g(\mu,\sigma) = \frac{USL-\mu}{3\sigma}.
\end{equation}

\paragraph{Step 1: Asymptotic covariance of $(\bar X,S)$.}

From Eq.~(\ref{eq:asymptotic_joint}), the vector $(\bar X,S)$ admits
an asymptotic normal distribution with covariance matrix $\Sigma$.
Under normality, this matrix has the explicit form
\[
\Sigma =
\begin{pmatrix}
	\sigma^2 & 0 \\
	0 & \sigma^2/2
\end{pmatrix}.
\]
The off-diagonal term is zero because $\bar X$ and $S^2$ are independent
for normal samples.

\paragraph{Step 2: Gradient of $g$.}
For the capability functional $g(\mu,\sigma)$ in Eq.~\eqref{cpku}, the gradient with respect to $(\mu,\sigma)$ is
\[
\nabla g(\mu,\sigma)
=
\begin{pmatrix}
	-\dfrac{1}{3\sigma} \\[6pt]
	-\dfrac{USL-\mu}{3\sigma^2}
\end{pmatrix}
=
\begin{pmatrix}
	-\dfrac{1}{3\sigma} \\[6pt]
	-\dfrac{C_{pk}^{true}}{\sigma}
\end{pmatrix}.
\]

\paragraph{Step 3: Asymptotic variance of $\widehat C_{pk}$.}
Substituting $\nabla g(\mu,\sigma)$ and $\Sigma$
into Eq.~(\ref{sigmac2}) gives
\[
\sigma_C^2
=
\left(\frac{1}{3\sigma}\right)^2 \sigma^2
+
\left(\frac{C_{pk}^{true}}{\sigma}\right)^2
\frac{\sigma^2}{2}.
\]

which simplifies to
\[
\sigma_C^2
=
\frac{1}{9}
+
\frac{(C_{pk}^{true})^2}{2},
\quad
\sigma_C
=
\sqrt{\frac{1}{9}+\frac{(C_{pk}^{true})^2}{2}}.
\]

\paragraph{Interpretation.}
The closed-form expression above provides a normal-theory
approximation for the $\sqrt{n}$-scale dispersion parameter
$\sigma_C$. It shows that boundary decision instability depends
only on the true capability level and not on the absolute scale
of $\sigma$. For example, when $C_{pk}^{true}=1.33$,
\[
\sigma_C
\approx
\sqrt{\frac{1}{9}+\frac{1.33^2}{2}}
\approx 1.05.
\]
Thus
\[
\Var(\widehat C_{pk})
\approx
\frac{\sigma_C^2}{n}
\approx
\frac{1.10}{n},
\]
so that the estimator has standard deviation approximately
$\sigma_C/\sqrt{n}$.

This magnitude explains why moderate sample sizes can still produce substantial
boundary crossing when $C_{pk}^{true}$ is near $C_0$.
In Section~\ref{sec4-d}, we further compare this normal-theory
approximation with Monte Carlo estimates obtained
under the exact plug-in estimator.

\medskip
\noindent
\textit{Remark (Non-regular case).}
If $C_{pu}^{true} = C_{pl}^{true}$, the minimum operator is non-differentiable at the true parameter. In this non-regular configuration, the delta-method expansion may fail, and the limiting acceptance probability need not converge to $1/2$ under local parameterization. This singular case corresponds to perfect centering under symmetric specifications and lies outside the regular regime characterized above. In such singular configurations, alternative asymptotic techniques (e.g., directional derivatives or empirical process methods) would be required, and the limiting decision probability may deviate from 1/2. In practice, exact symmetry is uncommon due to process drift, tool wear, and specification asymmetry, so the regular regime analyzed here captures the dominant operational setting encountered in manufacturing applications.

Extensions of the boundary-instability framework to non-normal capability definitions (e.g., percentile-based indices) are provided in Appendix~\ref{app:non_normal}.
\subsection{Finite-Sample Approximation of Misclassification Risk}
\label{sec3-d}

While Theorem~\ref{thm:boundary} characterizes the local asymptotic
behavior of the decision probability, finite-sample misclassification
risk depends on the exact sampling distribution of $\widehat{C}_{pk}$.
Because $\widehat{C}_{pk}$ is a nonlinear functional of $(\bar{X},S)$,
involving both the reciprocal of the sample standard deviation and
the nonsmooth $\min(\cdot)$ operator, closed-form expressions for
finite-sample threshold-crossing probabilities are generally unavailable.

We therefore estimate misclassification probabilities using Monte Carlo
simulation, which directly approximates the finite-sample behavior
of the decision rule.

Let $C_{pk}^{true}$ denote a fixed population capability level and
$n$ the sample size. Under normal sampling, data are generated from a
centered distribution with variance calibrated to achieve the target
$C_{pk}^{true}$. For each configuration $(C_{pk}^{true}, n)$,
we implement the following procedure.

\subsubsection{Monte Carlo estimation of misclassification risk}
For each configuration $(C_{pk}^{true}, n)$, misclassification probabilities are approximated by Monte Carlo simulation under the calibrated sampling model. In each experiment, the true capability level
$C_{pk}^{true}$ is fixed, and data are generated from a calibrated
sampling model whose parameters are chosen so that the process
attains the specified $C_{pk}^{true}$.

For $b = 1, \dots, B$, we generate an independent sample of size $n$,
compute the capability estimate $\widehat{C}_{pk}^{(b)}$, and record
the induced threshold decision
$$
D^{(b)} = \I\!\left(\widehat{C}_{pk}^{(b)} \ge C_0\right),
$$
where $D^{(b)} = 1$ denotes acceptance and $D^{(b)} = 0$ denotes rejection.

Because Type I and Type II misclassification probabilities are defined
under distinct parameter regimes, they are estimated in separate Monte
Carlo experiments:

\begin{itemize}
	\item Under the regime $C_{pk}^{true} < C_0$, the event
	$\{\widehat{C}_{pk}^{(b)} \ge C_0\}$ corresponds to a false accept.
	The Type I misclassification probability is therefore estimated by
	\begin{equation}
		\widehat{\Pr}_{\mathrm{Type\ I}}
		=
		\frac{1}{B}
		\sum_{b=1}^{B}
		\I\!\left(\widehat{C}_{pk}^{(b)} \ge C_0\right).
	\end{equation}
	
	\item Under the regime $C_{pk}^{true} \ge C_0$, the event
	$\{\widehat{C}_{pk}^{(b)} < C_0\}$ corresponds to a false reject.
	The Type II misclassification probability is therefore estimated by
	\begin{equation}
		\widehat{\Pr}_{\mathrm{Type\ II}}
		=
		\frac{1}{B}
		\sum_{b=1}^{B}
		\I\!\left(\widehat{C}_{pk}^{(b)} < C_0\right).
	\end{equation}
\end{itemize}
We define the general misclassification operator
\begin{equation}
	\mathcal{M}(C_{pk}^{true}, n, F, \mathcal{E})
	=
	\left(
	\widehat{\Pr}_{\mathrm{Type\ I}},
	\widehat{\Pr}_{\mathrm{Type\ II}}
	\right),
\end{equation}
which maps the true capability level $C_{pk}^{true}$, sample size $n$,
sampling distribution $F$, and estimator $\mathcal{E}$
to the induced decision risk under threshold-based approval.

In the present study, we consider the baseline configuration
consisting of the normal sampling model and the classical
$C_{pk}$ estimator. Denote this configuration by
$(F_0, \mathcal{E}_0)$. We therefore analyze the two-dimensional section
\begin{equation}
	M(C_{pk}^{true}, n)
	:=
	\mathcal{M}(C_{pk}^{true}, n, F_0, \mathcal{E}_0),
\end{equation}
which characterizes finite-sample decision instability as a function of the true capability level and sample size.

\subsubsection{Geometry of the Misclassification Surface and Local Scaling}
\begin{figure*}[!t]
	\centering
	\includegraphics[width=0.85\textwidth]{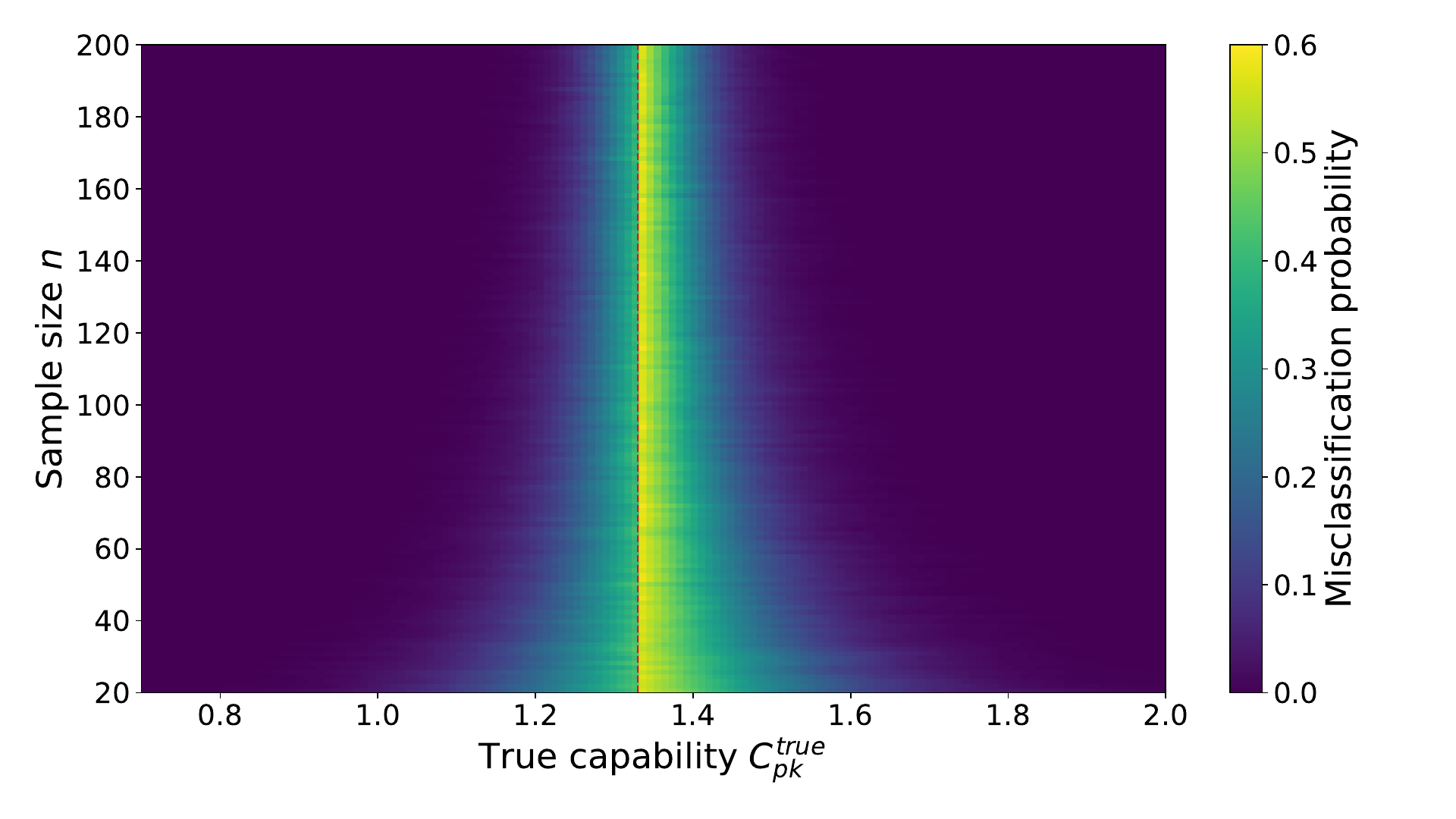}
	\caption{
		Finite-sample misclassification risk surface under normal sampling with bilateral specifications and threshold $C_0 = 1.33$. The horizontal axis denotes $C_{pk}^{true}$ and the vertical axis denotes sample size $n$. For $C_{pk}^{true}<C_0$, the plotted quantity is $\Pr(\widehat{C}_{pk}\ge C_0)$ (Type~I false accept); for $C_{pk}^{true}\ge C_0$, it is $\Pr(\widehat{C}_{pk}<C_0)$ (Type~II false reject). A pronounced ridge of elevated misclassification probability appears near $C_{pk}^{true}=C_0$.
	}
	\label{fig:misclass_surface}
\end{figure*}

Figure~\ref{fig:misclass_surface} reveals a ridge structure centered
at the decision boundary $C_{pk}^{true}=C_0$.
This geometry admits a direct analytical interpretation
through Theorem~\ref{thm:boundary}.
Under the local parameterization in Eq.~(\ref{cpkh}),
the acceptance probability converges to $\Phi(h/\sigma_C)$.
Hence the region in which misclassification probability
remains nondegenerate is precisely the $O(n^{-1/2})$
neighborhood of the boundary.

For example, when the sample size is $n$=32 (a typical engineering audit size) and the true capability lies within approximately $\pm0.05$ of the threshold, the misclassification probability exceeds 30\%. Even at $n$=50, substantial decision uncertainty remains near the boundary. These sample sizes are common in pilot production, dimensional audits, and supplier qualification, indicating that boundary instability is operationally realistic rather than purely theoretical.

The ridge arises when the distance between the true capability and the threshold is small relative to estimator dispersion. That is,
when the signal-to-noise ratio in Eq.~(\ref{eq:asymp_accept}),
\[
\left| \frac{\sqrt{n}\,(C_{pk}^{true}-C_0)}{\sigma_C} \right|,
\]
is close to zero. 
In this regime, the sampling distribution substantially overlaps the decision boundary, producing elevated misclassification risk. Because the signal-to-noise ratio is insufficient to separate the estimator from the threshold, moderate stochastic fluctuations in $(\bar{X},S)$ can reverse the pass/fail outcome.

As $n$ increases, estimator variance contracts at rate $1/n$,
and the width of the $O(n^{-1/2})$ neighborhood shrinks accordingly.
The ridge narrows, and misclassification probability decays,
consistent with the local asymptotic characterization of the theorem.

\begin{figure*}[!t]
	\centering
	\includegraphics[width=0.85\textwidth]{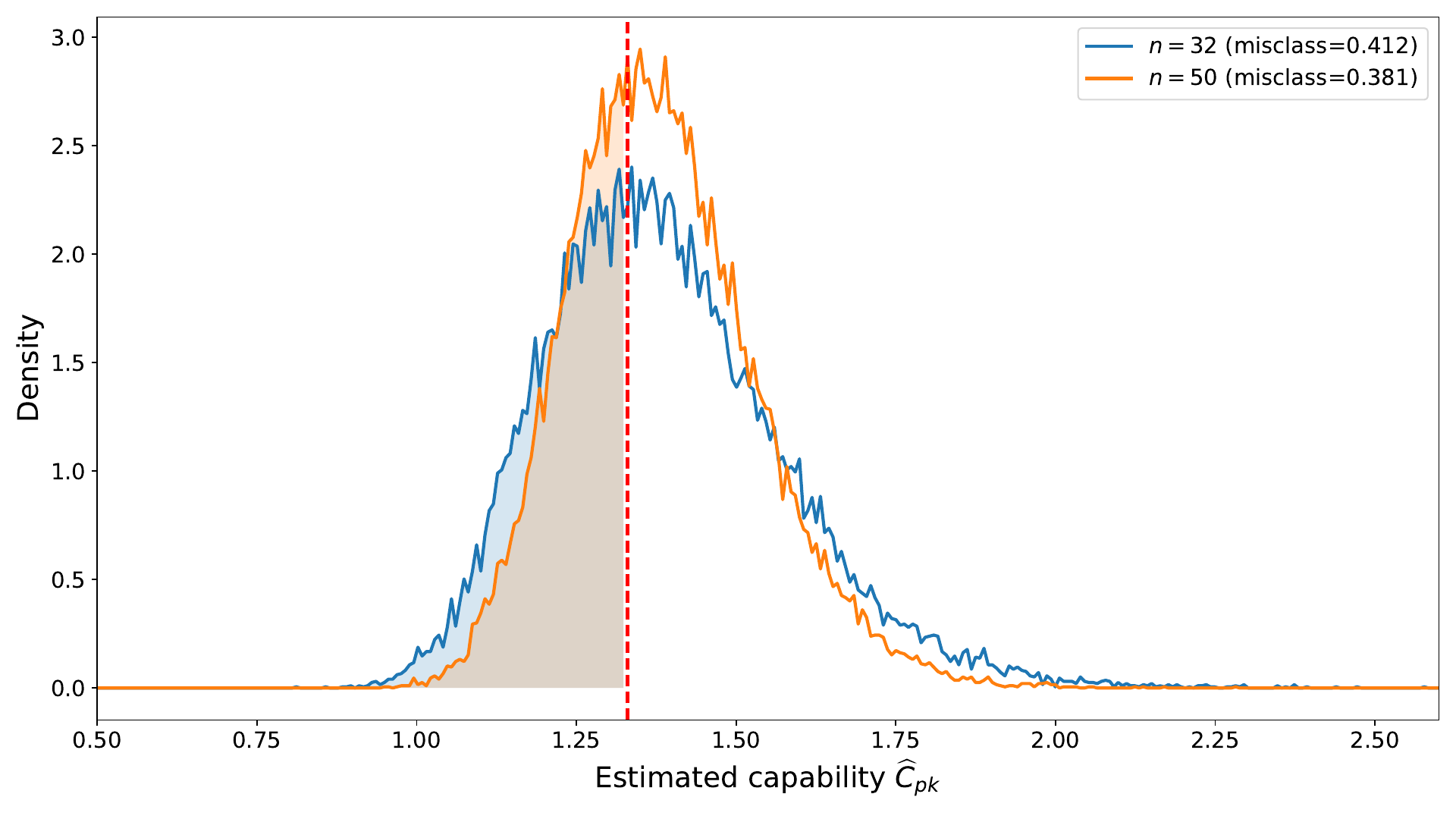}
	\caption{
		Sampling distributions of $\widehat{C}_{pk}$ near the threshold $C_0=1.33$.
		The dashed vertical line denotes the approval threshold.
		The shaded region represents the Type~II misclassification event
		$\{\widehat{C}_{pk} < C_0\}$ for a true-capable process
		($C_{pk}^{true}>C_0$).
		Two sample sizes are shown to illustrate variance contraction
		with increasing $n$. The legend reports the estimated misclassification probability.
	}
	\label{fig:cpk_sampling_near_threshold}
\end{figure*}

Figure~\ref{fig:cpk_sampling_near_threshold} provides a distributional
illustration of the same local asymptotic phenomenon.
For fixed $C_{pk}^{true}$, the estimator $\widehat{C}_{pk}$ has
a sampling distribution centered at $C_{pk}^{true}$
with dispersion proportional to $1/\sqrt{n}$.
Misclassification probability is exactly the probability mass
of this distribution falling on the opposite side of $C_0$.

When $C_{pk}^{true}$ lies in the $O(n^{-1/2})$ neighborhood
of $C_0$, the sampling distribution substantially overlaps
the boundary, producing elevated Type~I or Type~II risk.
Increasing $n$ tightens the distribution,
reducing the overlap and thereby shrinking
the decision-sensitive region.

Several systematic patterns emerge:

\begin{itemize}
	\item Misclassification risk increases sharply as sample size decreases.
	\item Risk is maximized in neighborhoods of the decision boundary.
	\item Even under ideal normal assumptions, moderate sample sizes
	can produce non-negligible decision instability.
\end{itemize}

\begin{figure*}[!t]
	\centering
	\subfloat[Scaling collapse under $\sqrt{n}$ local parameterization]{
		\label{fig3a}
		\includegraphics[width=0.475\textwidth]{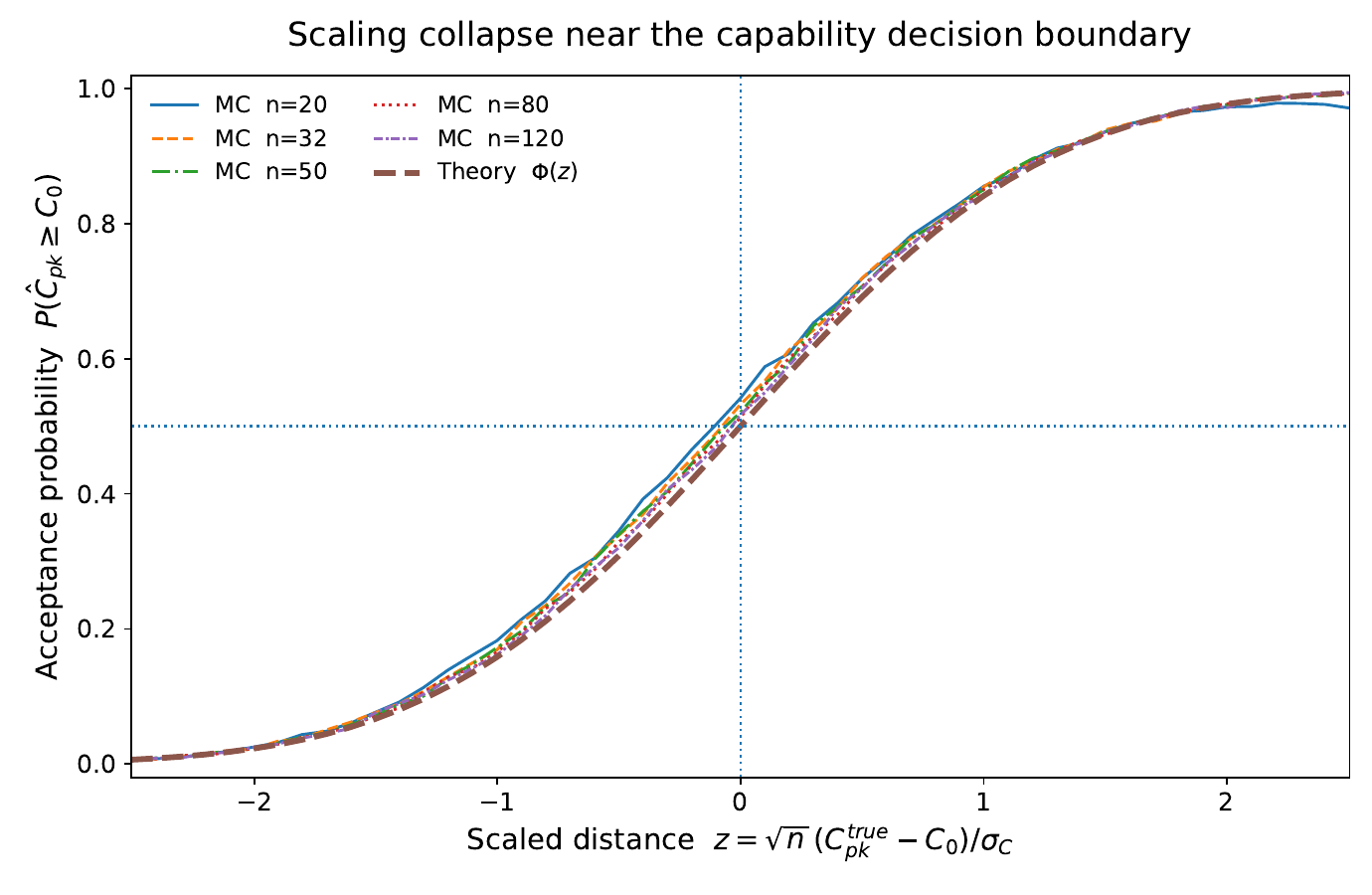}
	}
	\hspace{2pt}
	\subfloat[Residual: Monte Carlo minus theoretical $\Phi(z)$]{
		\label{fig3b}
		\includegraphics[width=0.475\textwidth]{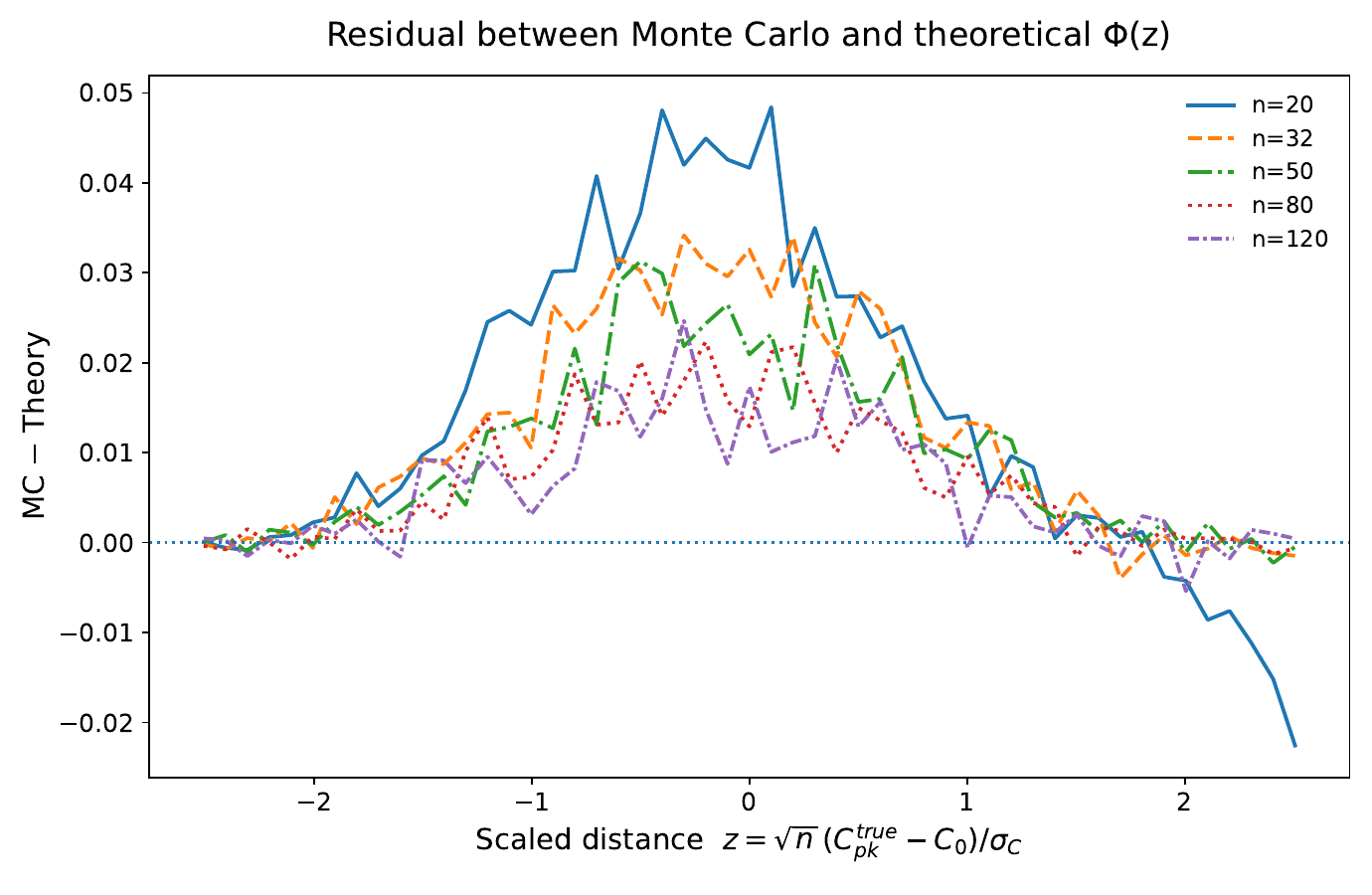}
	}
	\caption{
		Validation of $\sqrt{n}$ scaling near the capability decision boundary. Panel (a) shows the acceptance probability under the rescaled variable $z=\sqrt{n}(C_{pk}^{true}-C_0)/\sigma_C$. Monte Carlo curves for different sample sizes collapse and closely follow the theoretical prediction $\Phi(z)$, confirming that near the decision boundary classification behavior is governed by a single signal-to-noise parameter. The horizontal line at $0.5$ highlights the boundary instability at $C_{pk}^{true}=C_0$. Panel (b) displays the residual $\Delta(z)=P_{MC}(z)-\Phi(z)$. Deviations are small and decrease with sample size, demonstrating good finite-sample accuracy of the local asymptotic approximation.
	}
	\label{fig:collapse}
\end{figure*}

To further validate the local asymptotic characterization
in Theorem~\ref{thm:boundary},
we examine the $\sqrt{n}$ scaling structure
governing the decision probability.
Using the local parameterization introduced above,
and the corresponding rescaled signal-to-noise ratio $z = {\sqrt{n}\,(C_{pk}^{true}-C_0)}/{\sigma_C}$,
the theorem implies that
\[
\Pr(\widehat{C}_{pk} \ge C_0)
\;\rightarrow\;
\Phi(z),
\]
so that acceptance probability depends asymptotically
only on $z$\, rather than on $n$ and $C_{pk}^{true}$ separately.

Figure~\ref{fig:collapse}(a) reports Monte Carlo
acceptance probabilities under this scaling.
Curves corresponding to different sample sizes
collapse onto a common trajectory
and closely track the theoretical
$\Phi(z)$ prediction.
At $z=0$ (equivalently $C_{pk}^{true}=C_0$),
the acceptance probability approaches $0.5$,
confirming the boundary-instability result.

Figure~\ref{fig:collapse}(b) evaluates finite-sample accuracy
by plotting the residual
\[
\Delta(z)
=
P_{MC}(z) - \Phi(z),
\]
where $P_{MC}(z)$ denotes the Monte Carlo estimate.
The deviations are small in magnitude
and decrease with increasing sample size,
indicating that the $\sqrt{n}$ local approximation
provides an accurate quantitative description
of decision behavior in moderate samples.

The Monte Carlo analysis in this section approximates population-level misclassification probabilities under a calibrated sampling model. In contrast, the empirical analysis in Section~\ref{sec:val_instability} employs bootstrap resampling to quantify conditional decision instability given the observed dataset.

\begin{figure*}[!t]
	\centering
	\includegraphics[width=1.0\linewidth]{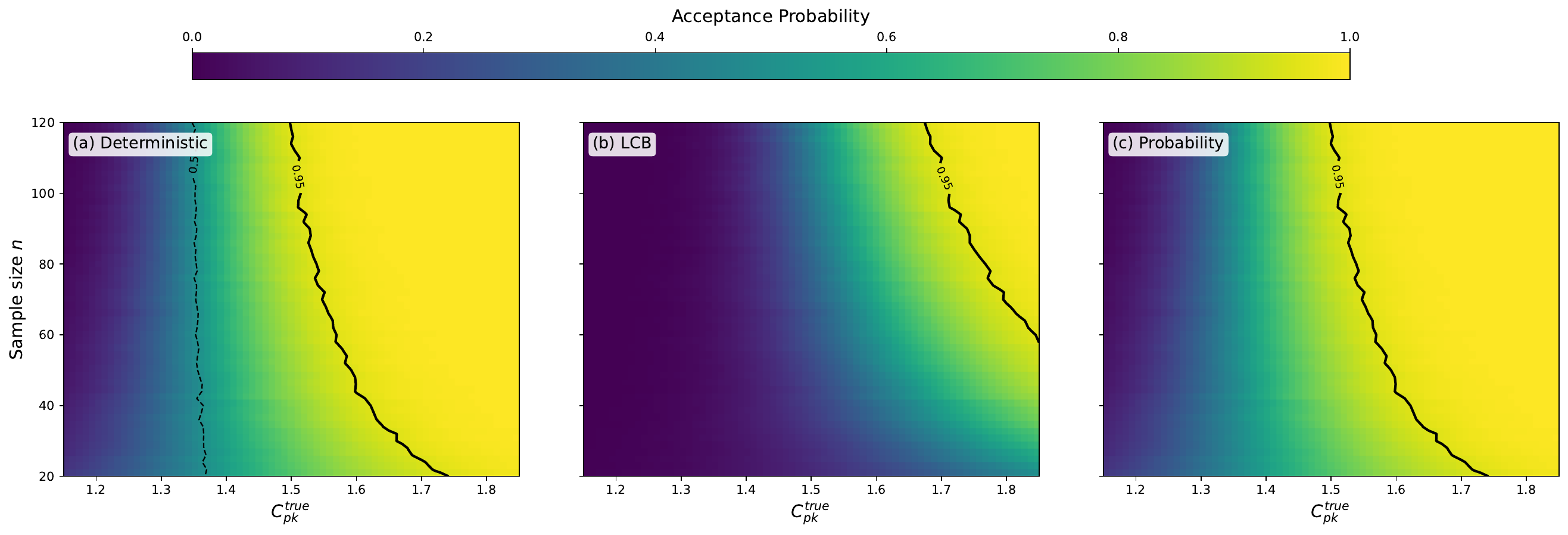}
	\caption{
		Acceptance probability surfaces under three capability approval rules as functions of the true population capability $C_{pk}^{true}$ and sample size $n$, with threshold $C_0 = 1.33$. Colors represent the probability of accepting the process under repeated sampling. (a) Deterministic rule: acceptance occurs when $\widehat{C}_{pk} \ge C_0$. The dashed contour denotes the $0.5$ acceptance boundary and the solid contour denotes the $0.95$ boundary. (b) Lower confidence bound (LCB) rule: acceptance requires $\mathrm{LCB}_{0.95} \ge C_0$. (c) Probability rule: acceptance requires $\Pr(\widehat{C}_{pk} \ge C_0) \ge 0.95$, estimated via Monte Carlo simulation.
	}
	\label{fig:decision_rules}
\end{figure*}

To illustrate how different capability approval rules behave under finite samples, Figure~\ref{fig:decision_rules} compares the acceptance probability surfaces induced by three representative decision rules: the classical deterministic threshold rule, a lower confidence bound (LCB) rule, and a probability-based acceptance rule.

Panel~(a) shows that under deterministic thresholding, the acceptance probability equals approximately $0.5$ when the true capability equals the threshold $C_0$, consistent with the boundary-instability result established in Theorem~\ref{thm:boundary}. Panels~(b) and~(c) illustrate how reliability-aware rules modify the effective approval boundary by incorporating estimator uncertainty. In particular, both the LCB and probability rules require higher true capability levels to achieve the same acceptance probability when the sample size is small, thereby reducing the risk of false acceptance. As the sample size increases, the acceptance boundaries under all rules converge toward the nominal threshold $C_{pk}^{true}=C_0$, reflecting diminishing sampling variability.

\section{Empirical Evidence of Near-Threshold Operating Regimes}
\label{sec4}

The boundary-instability analysis shows that threshold-based capability classification becomes intrinsically sensitive when the true capability lies in a neighborhood of the approval threshold $C_0$. This section examines whether such near-threshold operating regimes occur in practice.

We analyze a large-scale manufacturing dataset containing $N=880$ measured dimensions. Each dimension is sampled $n = 32$ times and is accompanied by a nominal value and bilateral tolerances (which may be asymmetric), inducing corresponding specification limits. Among the 880 dimensions, 582 pass a standard normality test at the chosen significance level, while the remaining dimensions exhibit detectable deviations from normality. For data protection, dimension-wise constant offsets were added to raw measurements and specification limits; these affine shifts preserve tolerance widths and do not affect capability indices or threshold comparisons.

The reported capability values are based on the conventional normal-assumption $C_{pk}$ formulation (see Eq. \ref{cpk2}) routinely used in production reporting. In this empirical study, the capability index is computed using the overall sample standard deviation, consistent with common industrial reporting practice. Because normal-based $C_{pk}$ reporting remains standard in many industrial environments even when mild non-normality is present, we retain the conventional estimator for all dimensions and additionally stratify results by normality-test outcome for robustness. For notational simplicity, we retain the generic notation $C_{pk}$ throughout.

For completeness, an explicit non-normal extension based on percentile-defined capability indices is outlined in Appendix~\ref{app:non_normal}.
\subsection{Concentration Near the Approval Threshold}
\label{sec:val_800_main}

Using the reported normal-based capability values, we examine empirical
concentration around the approval threshold $C_0 = 1.33$ for the full
dataset ($N=880$) and for the subset of $582$ dimensions that pass the
normality test.

\begin{table}[!t]
	\centering
	\renewcommand{\arraystretch}{1.0}
	\setlength{\tabcolsep}{6pt}
	\small
	\begin{tabular*}{\columnwidth}{ccc}
		\toprule
		Categories & $\widehat{C}_{pk} (N\!=\!880)$ & $\widehat{C}_{pk} (N\!=\!582)$ \\
		\midrule
		$|\widehat{C}_{pk} - 1.33| \le 0.0100$ & 4/880 (0.45\%) & 1/582 (0.17\%) \\
		$|\widehat{C}_{pk} - 1.33| \le 0.0200$ & 7/880 (0.80\%) & 3/582 (0.52\%) \\
		$|\widehat{C}_{pk} - 1.33| \le 0.0500$ & 18/880 (2.05\%) & 12/582 (2.06\%) \\
		$|\widehat{C}_{pk} - 1.33| \le 0.1000$ & 40/880 (4.55\%) & 25/582 (4.30\%) \\
		$|\widehat{C}_{pk} - 1.33| \le 0.1500$ & 51/880 (5.80\%) & 32/582 (5.50\%) \\
		$|\widehat{C}_{pk} - 1.33| \le 0.2000$ & 66/880 (7.50\%) & 44/582 (7.56\%) \\
		$|\widehat{C}_{pk} - 1.33| \le 0.2908$ & 100/880 (11.36\%) & 67/582 (11.51\%) \\
		\bottomrule
	\end{tabular*}
	\caption{Cumulative concentration of reported $\widehat{C}_{pk}$ values around the approval threshold $C_0 = 1.33$ for all $880$ dimensions and for the $582$ dimensions that pass the normality test. Percentages are shown in parentheses. The final band ($\pm 0.2908$) is derived from the scaled boundary $|z| = 1.645$ in Eq.~\eqref{eq:asymp_accept}. Under the local asymptotic formulation $z = \sqrt{n}(C_{pk}^{true} - C_0)/\sigma_C$, with $n = 32$ and $\sigma_C \approx 1$, this corresponds to $|C_{pk}^{true} - C_0| = 1.645\,\sigma_C/\sqrt{n} \approx 0.2908$. Dimensions within this band operate in a regime where the asymptotic one-sided boundary acceptance probability falls below approximately $95\%$.}
	\label{tab:threshold_concentration}
\end{table}

Table~\ref{tab:threshold_concentration} shows that concentration near
the threshold increases steadily as the tolerance band widens. While
fewer than 1\% of dimensions lie within $\pm 0.02$ of $1.33$,
approximately $4.6\%$ fall within $\pm 0.10$, and $7.5\%$ lie within
$\pm 0.20$. Extending to $\pm 0.2908$, more than $11\%$ of all
dimensions operate near the threshold.

The final band is directly motivated by the instability analysis in
Section~\ref{sec3}. Under the scaled formulation in Eq.~\eqref{eq:asymp_accept},
$|z| = 1.645$ implies $\Phi(1.645) \approx 0.95$.
Because the approval rule is one-sided ($\widehat{C}_{pk} \ge C_0$),
this corresponds to approximately a $5\%$ misclassification probability
for dimensions located at this boundary distance.
Thus, characteristics within $\pm 0.2908$ of the threshold operate in
a regime where asymptotic decision reliability falls below $95\%$.
The value $1.645$ is therefore appropriate for the one-sided
decision structure considered here; the conventional two-sided
$95\%$ cutoff ($|z|=1.96$) would correspond to a stricter $2.5\%$
tail probability and is less aligned with the threshold-crossing rule.

Importantly, the prevalence of near-threshold dimensions is nearly
identical for the full dataset and for the subset that passes the
normality test. The similarity across all bands indicates that
near-threshold concentration is not driven solely by non-normal
characteristics but is also present among approximately normal
dimensions.

Overall, a nontrivial fraction of dimensions operate in a narrow
neighborhood of the approval threshold. In view of the boundary
instability characterized in Section~\ref{sec3}, this empirical prevalence
suggests that a meaningful subset of real-world dimensions lie
precisely in the regime where threshold-based decisions are most
sensitive to sampling variability.

\subsection{Empirical Decision Instability via Resampling}
\label{sec:val_instability}

To quantify empirical decision instability in the observed data,
we conduct a resampling-based stability analysis for each of the
$880$ dimensions. For a given dimension $j$, let
$\widehat{C}_{pk,j}$ denote the reported normal-based capability index
computed from $n=32$ observations.

\paragraph{Bootstrap construction.}
For each dimension, we generate $B=5000$ nonparametric bootstrap
resamples of size $n$ (with replacement) from the original
measurements.
Let $\widehat{C}_{pk,j}^{*(b)}$ denote the conventional normal-based
capability estimate computed from the $b$-th bootstrap sample,
$b=1,\dots,B$.
Define the bootstrap approval indicator \cite{collins1995bootstrap}
\[
D_{j}^{*(b)}
=
\I\!\left(\widehat{C}_{pk,j}^{*(b)} \ge C_0\right),
\qquad C_0 = 1.33.
\]
The empirical approval frequency is estimated as
\[
\widehat{p}_j
=
\frac{1}{B}
\sum_{b=1}^{B}
D_{j}^{*(b)}.
\]
The decision flip rate is defined as
\[
\widehat{q}_j
=
\min\!\left(\widehat{p}_j,\,
1-\widehat{p}_j\right),
\]
which measures the probability that repeated sampling would
reverse the threshold-based approval outcome.
By construction, $\widehat{q}_j \in [0,0.5]$, attaining $0.5$
when approval and rejection occur with equal frequency
and equaling zero under perfectly stable decisions.

It is important to note that the bootstrap procedure estimates conditional decision instability given the observed dataset, rather than the population-level misclassification probability defined in Section~\ref{sec3-a}. The bootstrap therefore provides an empirical approximation to the stability of the threshold decision under repeated sampling from the observed process.

\paragraph{Distribution of instability.}
Across all $880$ dimensions, the median flip rate is $0$,
indicating that most characteristics are sufficiently distant
from the threshold to yield stable classifications.
However, instability is non-negligible for a meaningful subset.
Approximately $7.84\%$ of dimensions exhibit flip rates exceeding
$20\%$, and $4.77\%$ exceed $30\%$.
The 90th percentile of the flip-rate distribution is $0.0997$,
implying that the most unstable decile operates near
a $10\%$ decision-reversal probability.

\begin{figure}[htbp]
	\centering
	\includegraphics[width=1.00\linewidth]{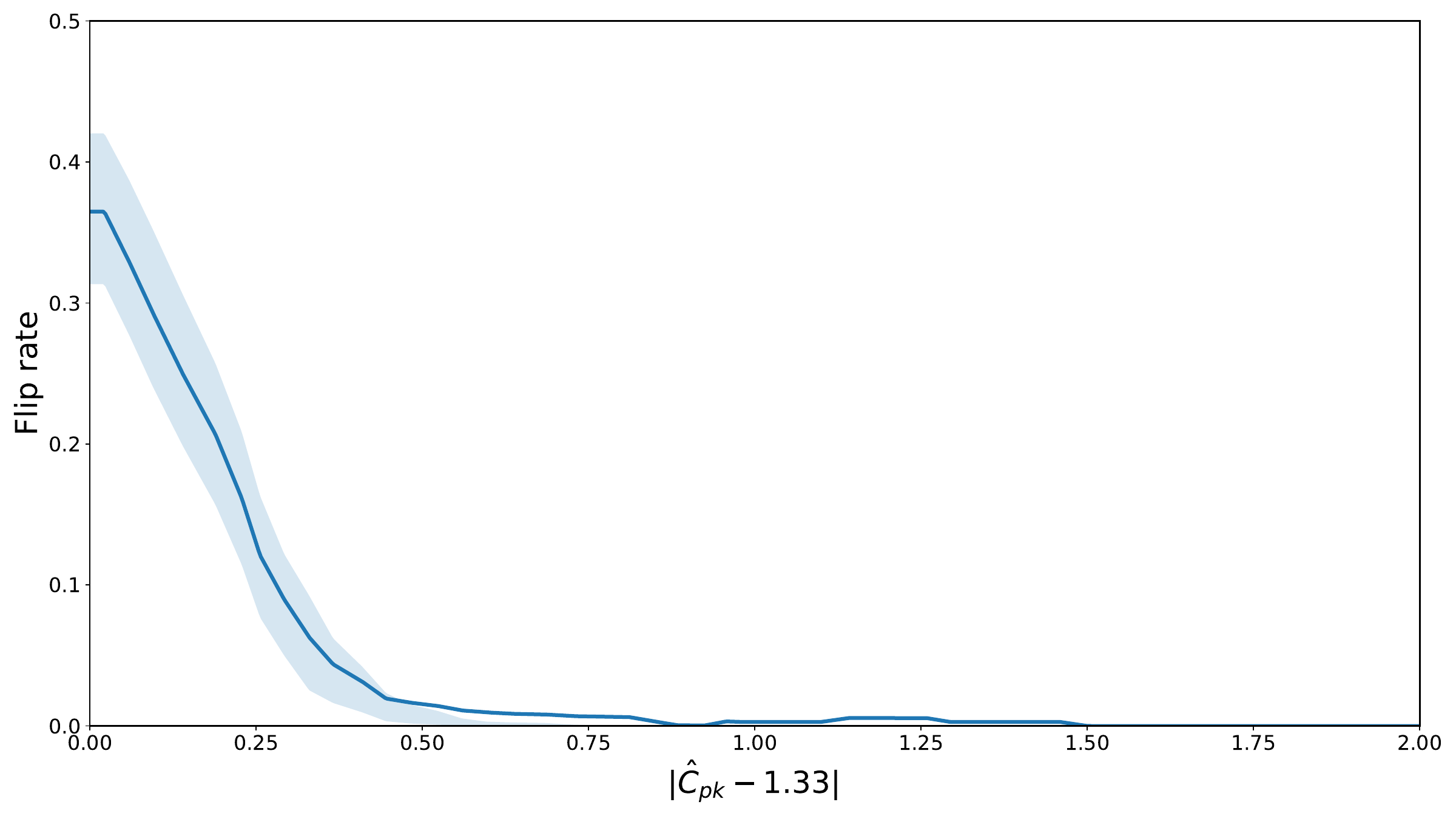}
	\caption{
		Empirical decision instability as a function of deterministic
		distance from the approval threshold.
		The solid curve represents the bin-averaged flip rate
		(using quantile-based bins), and the shaded band denotes
		the interquartile range within each bin.
		Bootstrap replication size: $B=5000$ per dimension.
	}
	\label{fig:instability_curve}
\end{figure}

\paragraph{Instability as a function of deterministic distance.}
Figure~\ref{fig:instability_curve} plots the bin-averaged flip rate
as a function of the deterministic distance
$|\widehat{C}_{pk,j} - C_0|$.
For visualization clarity, the horizontal axis is truncated at
$|\widehat{C}_{pk,j} - C_0| \le 2$, as dimensions beyond this range
exhibit near-zero flip probabilities and do not influence
the instability structure.

A pronounced instability ridge is observed near the approval
threshold.
Flip probabilities are largest when the estimated capability lies
close to $C_0 = 1.33$ and decay rapidly as the deterministic
distance increases.
Beyond approximately $0.5$ units from the threshold,
instability becomes negligible.
This empirical pattern aligns closely with the boundary-localized
misclassification structure derived in Section~\ref{sec3}.

\paragraph{Interpretation.}
These empirical findings closely mirror the boundary-instability
characterization developed in Section~\ref{sec3}.
Instability is not uniformly distributed across characteristics;
rather, it is sharply localized to a narrow neighborhood
of the approval threshold.
A nontrivial fraction of real manufacturing dimensions therefore
operate precisely in the regime where threshold-based capability
classification is most sensitive to sampling variability. 
The bootstrap approximates conditional decision instability given observed data, rather than population-level misclassification risk.

\subsection{Implication for Decision Stability}
\label{sec:val_implication}

Theorem~\ref{thm:boundary} establishes that decision behavior near the
approval threshold is governed by the signal-to-noise ratio
$\sqrt{n}(C_{pk}^{true}-C_0)/\sigma_C$.
When the true capability lies close to $C_0$, estimator dispersion
induces non-negligible probability mass on both sides of the threshold,
resulting in inherently unstable pass/fail classifications.

The empirical results in this section demonstrate that such
near-threshold regimes are not merely theoretical possibilities.
Approximately $11\%$ of dimensions lie within
$| \widehat C_{pk} - C_0 | \le 0.2908$,
and $7.84\%$ exhibit empirical flip probabilities exceeding $20\%$.
These findings indicate that boundary-sensitive operating conditions
are prevalent in manufacturing data.

Taken together, the theoretical characterization and empirical evidence
show that threshold-based capability approval is structurally fragile
for a nontrivial subset of dimensions.
Finite-sample variability can therefore materially influence release
decisions even when the underlying process remains unchanged.

\subsection{Design Implications for Threshold-Based Approval}
\label{sec4-d}
The boundary-instability result in Theorem~1 shows that
the conventional approval rule $\widehat{C}_{pk} \ge C_0$
induces a limiting acceptance probability of $1/2$ when
$C_{pk}^{true} = C_0$. Thus, fixed deterministic thresholds
leave boundary decision risk uncontrolled.

A natural modification is to introduce a 
$\sqrt{n}$-scaled safety margin. 
Such risk-controlled decision rules, often implemented through guard bands that shrink the effective acceptance region, are well established in conformity assessment standards ISO 14253-1:2013, where acceptance and rejection zones are explicitly adjusted to account for measurement uncertainty \cite{ISO14253-1-2013}.
\[
\widehat{C}_{pk} \ge C_0 + \frac{\kappa}{\sqrt{n}},
\]
where $\kappa > 0$ is chosen to control boundary acceptance risk. 
Such risk-adjusted decision rules are closely related in spirit to guard-band
approaches used in conformity assessment under uncertainty \cite{pendrill2014using}.

Under the local asymptotic expansion given in Eq.~\eqref{sigmac}, $\sqrt{n}(\widehat{C}_{pk} - C_{pk}^{true})$ converges in distribution to $\mathcal{N}(0,\sigma_C^2)$. The acceptance probability satisfies
\[
\Pr\!\left(
\widehat{C}_{pk} \ge C_0 + \tfrac{\kappa}{\sqrt{n}}
\right)
\approx
\Phi\!\left(
\frac{\sqrt{n}(C_{pk}^{true} - C_0) - \kappa}
{\sigma_C}
\right).
\]
At the boundary $C_{pk}^{true} = C_0$, this reduces to
\[
\Pr(\text{accept}\mid C_{pk}^{true}=C_0)
\approx
1 - \Phi\!\left(\frac{\kappa}{\sigma_C}\right).
\]
Therefore, selecting $\kappa = \sigma_C \Phi^{-1}(1-\alpha)$ ensures that the asymptotic boundary acceptance probability
is approximately $\alpha$.

\paragraph{Illustration.}
Under the baseline boundary configuration
($C_{pk}^{true}=1.33$, $n=32$), we estimate the $\sqrt{n}$-scale
dispersion parameter by Monte Carlo as
\[
\widehat{\sigma}_C
=
\mathrm{sd}\!\left(
\sqrt{n}(\widehat{C}_{pk}-C_{pk}^{true})
\right)
\approx 1.00,
\]
using the same plug-in estimator $\widehat{C}_{pk}$ as in
Eq.~\eqref{cpk2}. This estimate is close to the normal-theory
approximation derived in Section~\ref{sec3-c}
($\sigma_C \approx 1.05$ at $C_{pk}^{true}=1.33$),
and we adopt the simulation-based value here to ensure
full consistency with the finite-sample estimator and
sampling design used throughout the empirical analysis.

Equivalently, this corresponds to
\[
\mathrm{SD}(\widehat{C}_{pk})
\approx
\frac{\widehat{\sigma}_C}{\sqrt{n}}
\approx
\frac{1.00}{\sqrt{32}}
\approx 0.177,
\]
which reflects the practical magnitude of estimator
dispersion at typical audit sample sizes.

For a target boundary acceptance risk $\alpha=0.05$,
\[
\kappa
=
\sigma_C \Phi^{-1}(1-\alpha)
\approx
1.00 \times 1.645
\approx 1.645,
\]
yielding a safety margin
\[
\frac{\kappa}{\sqrt{32}}
=
\frac{1.645}{5.657}
\approx 0.2908.
\]
The adjusted approval threshold therefore becomes
\[
\widehat{C}_{pk} \ge 1.33 + 0.2908 = 1.621.
\]

This margin decreases at rate $1/\sqrt{n}$ as sample size
increases, preserves the original capability definition,
and converts the uncontrolled $50\%$ boundary acceptance
probability into a calibrated $\alpha$-level boundary
acceptance rate.

\paragraph{Practical interpretation.} The above calibration should not be interpreted as a recommendation to universally replace the conventional threshold by a substantially larger fixed threshold. Rather, it quantifies the implicit boundary risk embedded in deterministic approval rules under moderate sample sizes. The resulting margin illustrates the scale of sampling-induced decision volatility when operating near the approval boundary. In practice, organizations may instead adopt alternative risk-adjusted strategies, such as lower confidence bounds or context-specific guard bands, depending on contractual, economic, or regulatory considerations. The present analysis provides a quantitative reference for evaluating such design choices.
\section{Conclusion}

This paper reframes threshold-based process capability approval as a finite-sample statistical decision problem rather than a deterministic comparison of estimated indices with fixed thresholds. Although indices such as $C_{pk}$ are commonly
interpreted relative to fixed thresholds, we show that under realistic
sample sizes, threshold-based approval is inherently probabilistic.

By formalizing approval as the stochastic decision rule in Eq. \eqref{eq:decision_rule}, we characterized decision reliability through
misclassification probabilities. Under standard regularity conditions,
we established a boundary-instability result demonstrating that when
$C_{pk}^{true}=C_0$, the acceptance probability converges to $0.5$.
Thus deterministic thresholding cannot yield stable decisions when the
true process capability lies near the approval boundary.

Monte Carlo simulations further revealed that finite-sample
misclassification risk concentrates in a ridge near the threshold and
decays at the canonical $1/\sqrt{n}$ rate as sample size increases.
Empirical analysis of an 880-dimension manufacturing dataset confirms
that such near-threshold operating regimes are common in practice,
indicating that boundary-sensitive decisions are operationally relevant.

Overall, the analysis provides a statistical foundation for interpreting
threshold-based capability approval under sampling uncertainty.
Because smooth $\sqrt{n}$-consistent estimators are combined with
deterministic classification rules, approval decisions near the
threshold inevitably exhibit stochastic volatility. For moderate
sample sizes ($n\approx20$-$50$), processes operating within
$\pm0.05$ of the approval threshold may exhibit misclassification
probabilities exceeding $30\%$.

These findings suggest that approval reliability depends not only on
estimator accuracy but also on the interaction between estimator
dispersion and fixed thresholds. Risk-adjusted approval margins or
confidence-bound-based criteria may therefore provide more stable
decision rules, representing a promising direction for future research.

\appendix
\numberwithin{equation}{section}
\numberwithin{figure}{section}
\section*{APPENDIX}
\section{Extension to Non-Normal Capability Models}
\label{app:non_normal}

The main text develops the boundary-instability result under
classical normal-based capability estimation.
We briefly outline how the decision-theoretic framework
extends to percentile-based capability definitions
commonly used under non-normal sampling.

We denote the percentile-based capability index by $C_{Npk}$ 
to distinguish it from the classical normal-based $C_{pk}$. 
This notation reflects the underlying distributional specification 
rather than a fundamentally different decision functional; 
the same threshold-based classification framework applies.

\subsection{Percentile-Based Capability Index}

When the normality assumption is violated,
capability is often defined through distribution percentiles
to better reflect empirical tail behavior.
Let $P_{0.135}$, $P_{50}$, and $P_{99.865}$
denote the lower 0.135th percentile,
median, and upper 99.865th percentile
of the underlying distribution.
For bilateral specifications,
the percentile-based capability index is defined as \cite{chen1997application, chen2001new, anis2008basica, kovarik2014process}

\begin{equation}
	C_{Npk}
	=
	\min\!\left(
	\frac{\USL - P_{50}}{P_{99.865} - P_{50}},
	\frac{P_{50} - \LSL}{P_{50} - P_{0.135}}
	\right).
	\label{eq:cnpk_appendix}
\end{equation}

When the underlying distribution is normal,
this formulation reduces algebraically
to the classical $C_{pk}$.

\subsection{Asymptotic Behavior Under Regularity}
\begin{figure*}[!t]
	\centering
	\includegraphics[width=0.85\textwidth]{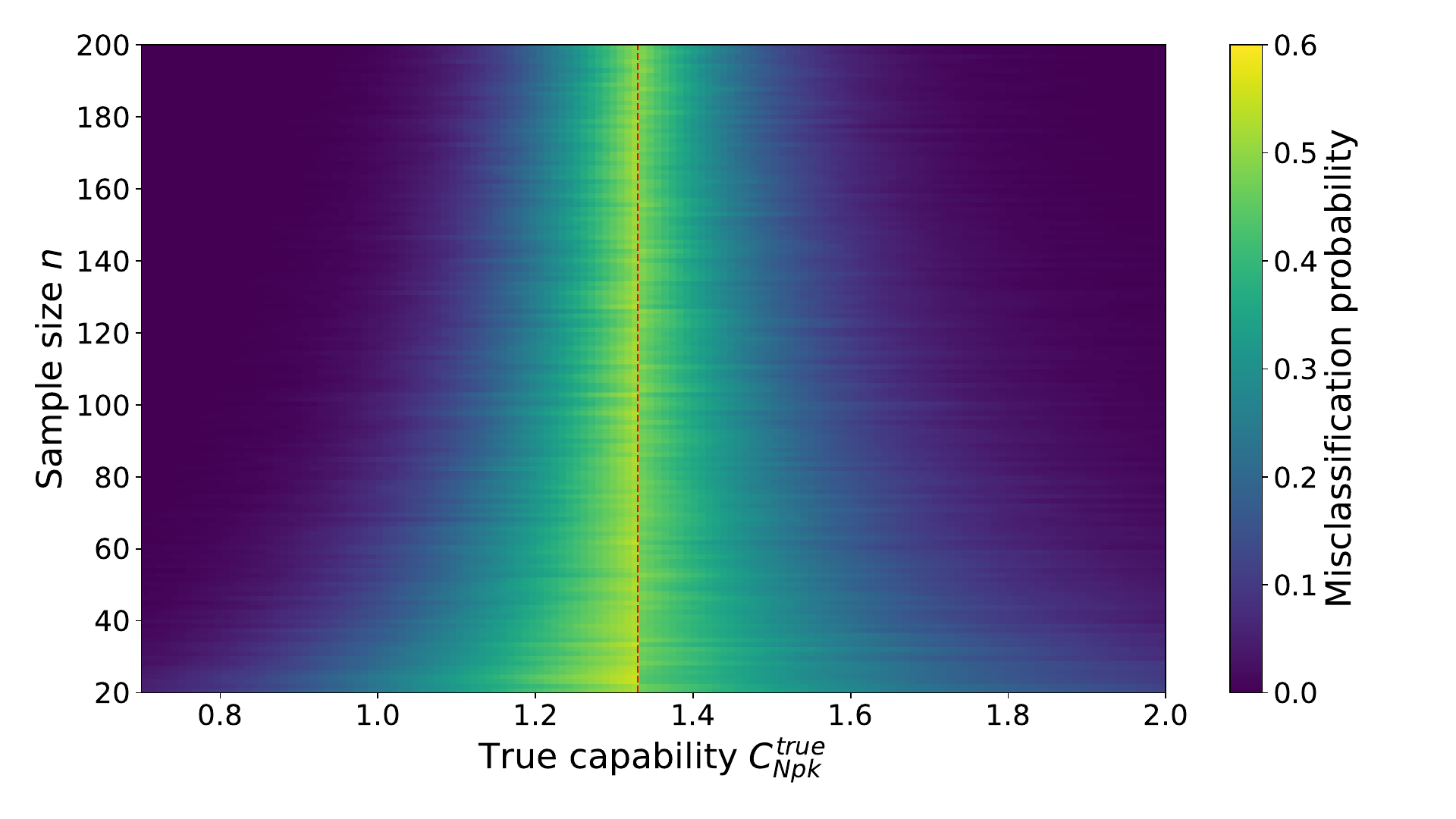}
	\caption{
		Finite-sample misclassification risk surface under a shifted
		lognormal process using the percentile-based estimator
		$\widehat{C}_{Npk}$.
		The horizontal axis denotes the true capability
		$C_{Npk}^{true}$ and the vertical axis denotes
		sample size $n$.
		Instability remains concentrated near the threshold,
		although skewness alters the shape and asymmetry
		of the ridge relative to the normal case.
	}
	\label{fig:misclass_surface_lognormal}
\end{figure*}
Under standard regularity conditions (e.g., the underlying distribution admits a continuous density that is positive in a neighborhood of $P_\alpha$), sample quantile estimators are asymptotically normal. These regularity conditions are standard in quantile asymptotics and are satisfied for a wide class of continuous distributions, including lognormal, gamma, and Weibull models commonly used in reliability engineering. Specifically, if $\widehat{P}_\alpha$ denotes the empirical or parametric estimator of the $\alpha$-quantile, then
\[
\sqrt{n}\left(
\widehat{P}_\alpha - P_\alpha
\right)
\;\Rightarrow\;
\mathrm{N}(0,\tau_\alpha^2),
\]
where $\tau_\alpha^2$ depends on the density
of the underlying distribution at $P_\alpha$.

Provided that the minimum operator in
\eqref{eq:cnpk_appendix}
is uniquely attained at the true parameter,
the capability functional remains locally differentiable.
Applying the multivariate delta method yields

\begin{equation}
	\sqrt{n}\left(
	\widehat{C}_{Npk}
	-
	C_{Npk}^{true}
	\right)
	\Rightarrow
	\mathrm{N}(0,\sigma_{CN}^2).
\end{equation}
Here $\sigma_{CN}^2 = \nabla g^\top \Sigma_Q \nabla g$,
where $g$ denotes the percentile-based capability functional
and $\Sigma_Q$ is the asymptotic covariance matrix of the 
vector of sample quantile estimators.

Consequently, the acceptance probability under
the threshold rule
\begin{equation}
	D = \I(\widehat{C}_{Npk} \ge C_0)
\end{equation}
admits the same local scaling structure as in
Theorem~\ref{thm:boundary} derived for the normal-based estimator.

In particular, the local parameterization in Eq.~\eqref{cpkh} applies here with $C_{pk}$ replaced by $C_{Npk}$. We obtain
\begin{equation}
	\Pr(\widehat{C}_{Npk} \ge C_0)
	\rightarrow
	\Phi\!\left(\frac{h}{\sigma_{CN}}\right).
\end{equation}

Thus, boundary instability is not confined
to the normal model; it arises generically
whenever the capability estimator
admits a regular $\sqrt{n}$ asymptotic expansion.

The boundary-instability mechanism extends directly to target-based capability indices, since these estimators also admit $\sqrt{n}$-consistent asymptotic expansions under regularity conditions.

\subsection{Illustrative Finite-Sample Behavior Under Lognormal Sampling}

To illustrate finite-sample behavior under asymmetric distributions, we consider a shifted lognormal process calibrated to achieve a specified true capability $C_{Npk}^{true}$. Monte Carlo simulation is used to approximate misclassification probabilities induced by the threshold rule $\I(\widehat{C}_{Npk} \ge C_0)$. The resulting misclassification surface is shown in Figure~\ref{fig:misclass_surface_lognormal}.

The ridge structure observed under normal sampling persists under non-normal conditions. Distributional asymmetry modifies the width and symmetry of the instability region through changes in estimator dispersion, but the local signal-to-noise mechanism governing decision instability remains intact.

\section*{DECLARATION}
\begin{description}
	\item[Funding:] This study received no external funding.
	
	\item[Conflicts of interest:] The authors declare no conflicts of interest.
	
	\item[Availability of data and material:] The empirical dataset is derived from anonymized modified manufacturing data. Processed data are available from the corresponding author upon reasonable request.
	
	\item[Code availability:] Simulation and analysis code are available from the corresponding author upon reasonable request.
	
	\item[Ethics approval:] Not applicable.
	
	\item[Consent for publication:] All authors approve the final manuscript.
\end{description}
\bibliographystyle{unsrtnat}
\bibliography{references}

@article{anis2008basica,
  title = {Basic {{Process Capability Indices}}: {{An Expository Review}}},
  shorttitle = {Basic {{Process Capability Indices}}},
  author = {Anis, Mohammed Z.},
  year = {2008},
  month = dec,
  journal = {International Statistical Review},
  volume = {76},
  number = {3},
  pages = {347--367},
  issn = {0306-7734, 1751-5823},
  doi = {10.1111/j.1751-5823.2008.00060.x},
  urldate = {2024-12-13}
}

@article{chen1997application,
  title = {An Application of Non-Normal Process Capability Indices},
  author = {Chen, K. S. and Pearn, W. L.},
  year = {1997},
  journal = {Quality and Reliability Engineering International},
  volume = {13},
  number = {6},
  pages = {355--360},
  issn = {1099-1638},
  doi = {10.1002/(SICI)1099-1638(199711/12)13:6<355::AID-QRE125>3.0.CO;2-V},
  urldate = {2024-12-13},
  copyright = {Copyright {\copyright} 1997 John Wiley \& Sons, Ltd.}
}

@article{chen2001new,
  title = {A New Process Capability Index for Non-normal Distributions},
  author = {Chen, Jann-Pygn and Ding, Cherng G.},
  year = {2001},
  month = oct,
  journal = {International Journal of Quality \& Reliability Management},
  volume = {18},
  number = {7},
  pages = {762--770},
  issn = {0265-671X},
  doi = {10.1108/02656710110396076},
  urldate = {2024-12-13},
  copyright = {https://www.emerald.com/insight/site-policies}
}

@article{kane1986process,
  title = {Process {{Capability Indices}}},
  author = {Kane, Victor E.},
  year = {1986},
  month = jan,
  journal = {Journal of Quality Technology},
  volume = {18},
  number = {1},
  pages = {41--52},
  publisher = {Taylor \& Francis},
  issn = {0022-4065},
  doi = {10.1080/00224065.1986.11978984},
  urldate = {2025-02-20}
}

@article{kotz2002process,
  title = {Process {{Capability Indices}}---{{A Review}}, 1992--2000},
  author = {Kotz, Samuel and Johnson, Norman L.},
  year = {2002},
  month = jan,
  journal = {Journal of Quality Technology},
  volume = {34},
  number = {1},
  pages = {2--19},
  issn = {0022-4065, 2575-6230},
  doi = {10.1080/00224065.2002.11980119},
  urldate = {2024-12-13}
}

@article{kovarik2014process,
  title = {Process {{Capability Indices}} for {{Non-Normal Data}}},
  author = {Kov{\'a}{\v r}{\'i}k, Martin and Sarga, Libor},
  year = {2014},
  volume = {11}
}

@article{pearn1992distributional,
  title = {Distributional and {{Inferential Properties}} of {{Process Capability Indices}}},
  author = {Pearn, W. L. and Kotz, Samuel and Johnson, Norman L.},
  year = {1992},
  month = oct,
  journal = {Journal of Quality Technology},
  volume = {24},
  number = {4},
  pages = {216--231},
  issn = {0022-4065, 2575-6230},
  doi = {10.1080/00224065.1992.11979403},
  urldate = {2024-12-13}
}

@book{juran1979quality,
	title={Quality control handbook},
	author={Juran, Joseph M and Gryna, Frank M and Bingham, Richard S},
	volume={3},
	year={1979},
	publisher={McGraw-hill New York}
}

@article{ISO22514-4-2016,
	title={Statistical methods in process management -- Capability and performance -- Part 4: Process capability estimates and performance measures},
	author={{ISO/TR}},
	journal={},
	volume={},
	pages={},
	year={ISO/TR 22514-4:2016 (2016)},
	publisher={}
}

@article{ISO11462-1-2001,
	title={Guidelines for implementation of statistical process control (SPC) - Part 1: Elements of SPC},
	author={ISO},
	journal={},
	volume={},
	pages={},
	year={ISO 11462-1:2001 (2001)},
	publisher={}
}

@article{ISO11462-2-2010,
	title={Guidelines for implementation of statistical process control (SPC) - Part 2: Catalogue of tools and techniques},
	author={{ISO}},
	journal={},
	volume={},
	pages={},
	year={ISO 11462-2:2010 (2010)},
	publisher={}
}

@book{van2000asymptotic,
	title={Asymptotic statistics},
	author={Van der Vaart, Aad W},
	volume={3},
	year={2000},
	publisher={Cambridge university press}
}

@article{zappa2009misclassification,
	title={Misclassification rates, critical values and size of the design in measurement systems capability studies},
	author={Zappa, Diego and Deldossi, Laura},
	journal={Applied Stochastic Models in Business and Industry},
	volume={25},
	number={5},
	pages={601--611},
	year={2009},
	publisher={Wiley Online Library}
}

@article{burdick2005confidence,
	title={Confidence intervals for misclassification rates in a gauge R\&R study},
	author={Burdick, Richard K and Park, You-Jin and Montgomery, Douglas C and Borror, Connie M},
	journal={Journal of quality technology},
	volume={37},
	number={4},
	pages={294--303},
	year={2005},
	publisher={Taylor \& Francis}
}

@article{kushler1992confidence,
	title={Confidence bounds for capability indices},
	author={Kushler, Robert H and Hurley, Paul},
	journal={Journal of Quality Technology},
	volume={24},
	number={4},
	pages={188--195},
	year={1992},
	publisher={Taylor \& Francis}
}

@article{grau2011lower,
	title={Lower confidence bound for capability indices with asymmetric tolerances and gauge measurement errors},
	author={Grau, Daniel},
	journal={International Journal of Quality Engineering and Technology},
	volume={2},
	number={3},
	pages={212--228},
	year={2011},
	publisher={Inderscience Publishers}
}

@article{mathew2007generalized,
	title={Generalized confidence intervals for process capability indices},
	author={Mathew, Thomas and Sebastian, George and Kurian, KM},
	journal={Quality and reliability engineering international},
	volume={23},
	number={4},
	pages={471--481},
	year={2007},
	publisher={Wiley Online Library}
}

@book{lehmann1998theory,
	title={Theory of point estimation},
	author={Lehmann, Erich Leo and Casella, George},
	year={1998},
	publisher={Springer}
}

@article{zhang1990interval,
	title={Interval estimation of process capability index Cpk},
	author={Zhang, NF and Stenback, GA and Wardrop, DM},
	journal={Communications in Statistics-Theory and Methods},
	volume={19},
	number={12},
	pages={4455--4470},
	year={1990},
	publisher={Taylor \& Francis}
}

@article{collins1995bootstrap,
	title={Bootstrap confidence limits on process capability indices},
	author={Collins, Alan J},
	journal={Journal of the Royal Statistical Society: Series D (The Statistician)},
	volume={44},
	number={3},
	pages={373--378},
	year={1995},
	publisher={Wiley Online Library}
}

@article{pearn2004testing,
	title={Testing process performance based on capability index Cpk with critical values},
	author={Pearn, Wen Lea and Lin, PC},
	journal={Computers \& Industrial Engineering},
	volume={47},
	number={4},
	pages={351--369},
	year={2004},
	publisher={Elsevier}
}

@article{pendrill2014using,
	title={Using measurement uncertainty in decision-making and conformity assessment},
	author={Pendrill, Leslie R},
	journal={Metrologia},
	volume={51},
	number={4},
	pages={S206--S218},
	year={2014},
	publisher={IOP Publishing}
}

@article{ISO14253-1-2013,
	title={Geometrical product specifications (GPS) -- Inspection by measurement of workpieces and measuring equipment -- Part 1: Decision rules for proving conformity or nonconformity with specifications},
	author={ISO},
	journal={International Organization for Standardization},
	volume={},
	pages={},
	year={ISO 14253-1:2013 (2013)},
	publisher={ISO}
}

@book{berger2013statistical,
	title={Statistical decision theory and Bayesian analysis},
	author={Berger, James O},
	year={2013},
	publisher={Springer Science \& Business Media}
}

@book{casella2024statistical,
	title={Statistical inference},
	author={Casella, George and Berger, Roger},
	year={2024},
	publisher={Chapman and Hall/CRC}
}

@article{jiang2026practical,
	author  = {Jiang, Fei and Yang, Lei},
	title   = {Practical process capability indices workflows},
	journal = {The International Journal of Advanced Manufacturing Technology},
	year    = {2026},
	doi     = {10.1007/s00170-026-17782-7},
	url     = {https://doi.org/10.1007/s00170-026-17782-7}
}
\end{document}